\documentclass{article}
\usepackage{spconf,amsmath,graphicx}
\usepackage{xcolor}
\usepackage{cite}
\usepackage{booktabs}
\usepackage{multirow}
\usepackage{amsfonts}
\usepackage{comment}
\usepackage{float}
\usepackage[hidelinks]{hyperref}

\usepackage{enumitem}
\setlist{nosep, leftmargin=14pt}

\usepackage{mwe} 

\title{Exploring Foundation Models Fine-Tuning for Cytology Classification}

\name{
\parbox{\linewidth}{\centering
Manon Dausort$^{\star,1}$ \hspace{3mm} Tiffanie Godelaine$^{\star,1}$ \hspace{3mm} Maxime Zanella$^{1,2}$ \hspace{3mm} Karim El Khoury$^{1}$ \\Isabelle Salmon$^{3}$ \hspace{4mm} Beno\^it Macq$^{1}$\thanks{$^\star$The authors have contributed equally to this work.}}}

\address{$^{1}$ ICTEAM, Université Catholique de Louvain, Belgium\\ $^{2}$ ILIA, Université de Mons, Belgium\\ $^{3}$ CMMI, Université Libre de Bruxelles, Belgium}

\begin{document}
%

\maketitle

\begin{abstract}
Cytology slides are essential tools in diagnosing and staging cancer, but their analysis is time-consuming and costly. Foundation models have shown great potential to assist in these tasks. In this paper, we explore how existing foundation models can be applied to cytological classification. More particularly, we focus on low-rank adaptation, a parameter-efficient fine-tuning method suited to few-shot learning. We evaluated five foundation models across four cytological classification datasets. Our results demonstrate that fine-tuning the pre-trained backbones with LoRA significantly improves model performance compared to fine-tuning only the classifier head, achieving state-of-the-art results on both simple and complex classification tasks while requiring fewer data samples. Our source code is available on GitHub \href{https://github.com/mdausort/Cytology-fine-tuning}{https://github.com/mdausort/Cytology-fine-tuning}.
\end{abstract}

\begin{keywords}
Cytology classification, Parameter-efficient Fine-tuning, Low-rank Adaptation, Transfer Learning, Few-shot Learning
\end{keywords}

\section{Introduction}
\label{sec:intro}
Cytology slides are vital for diagnosing and staging cancer, offering detailed views of abnormal cells to guide treatment decisions~\cite{shaw2000history}. Analyzing these slides is labor-
intensive and costly, leading to delays in the reporting process, making automation essential to improve both classification efficiency and accuracy~\cite{JIANG2023102691}.

\vspace{2mm}

An effective approach to automate classification involves using deep learning models. However, the performance of these models is highly dependent on the quantity and diversity of data~\cite{Gong_2019}. Since handling large-scale cytology images demands high computational resources, a patch-based method is often chosen to divide gigapixel slides into thousands of smaller patches suitable for model training~\cite{Zhang_2017}. Annotating thousands of patches is cumbersome and challenging, as it must be performed by trained medical practitioners and requires a considerable amount of time~\cite{JIANG2023102691}. As a result, the community is exploring effective ways to utilize limited labeled data, an area of research known as few-shot learning~\cite{song2022fsl}

\vspace{2mm}

In parallel, large-scale models often referred to as foundation models (FMs) are being developed through extensive pre-training on large datasets~\cite{Radford2021clip, wu2020visual} and are valued for their robust generalization capabilities, especially in few-shot regimes. They have the potential to address data scarcity in more specific downstream tasks. Hence, the medical imaging community has been gathering large datasets to exploit the potential of foundational models, leading to medical imaging-oriented FMs~\cite{Zhang2024biomedclip, huang2023plip, Oluchi2023quilt, Lu2024conch,Chen2024uni} and paving the way for further research extensions in the field~\cite{zanella2024boostingvlms,ding2024}. In the context of digital pathology, histology has garnered considerable attention leading to the introduction of several FMs specifically designed for histological analysis~\cite{huang2023plip, Oluchi2023quilt, Lu2024conch, Chen2024uni}. For instance, UNI~\cite{Chen2024uni} was pre-trained on an extensive dataset comprising over 100 million images across 20 tissue types, totaling 77 TB of data. In contrast, cytology remains underexplored, largely due to the scarcity of large publicly available datasets. In comparison, the most extensive cytology dataset available to the public is composed of 40,229 images yet it only covers a single pathology~\cite{cai2024hicervix}. Due to the extensive data, time, and energy consumption required for training those FMs, we are encouraged to consider fine-tuning existing FMs as an efficient solution for cytology classification.

\vspace{2mm}

Traditionally, fine-tuning these models relies on computationally intensive methods. However, recent parameter-efficient fine-tuning approaches reduce the number of trainable parameters, enhancing generalization with minimal labeled data~\cite{zaken2022bitfits, hu2021lora}. Among these methods, Low Rank Adaptation (LoRA)~\cite{hu2021lora} allows model training without increasing the number of parameters at inference and has shown strong potential in low-data scenarios across various medical tasks and modalities~\cite{dutt2023parameter}. Motivated by the potential of LoRA demonstrated in previous studies, we aim to extend its application to cytology classification.

\vspace{2mm}

In this work, we investigate the applicability of existing FMs for cytology tasks through three experiments. First, we assess whether domain-specific models offer greater adaptability than general-purpose ones and explore the potential for transferring knowledge from histology to cytology. Then, we examine the benefits of using LoRA fine-tuning in few-shot scenarios. Finally, we fine-tune the most promising vision encoder with increasing amount of data to achieve state-of-the-art performance. We evaluate five FMs pre-trained on natural, biomedical and histopathological images on four cytology classification benchmark datasets and \textbf{observe that}:

\vspace{2mm}

\begin{itemize}
    \item Histology-specific models show superior adaptability as feature extractors for cytology classification.

    \item General models achieve better generalization when their weights are fine-tuned with LoRA in few-shot settings.

    \item CLIP's vision encoder fine-tuned with LoRA achieves state-of-the-art performance on a challenging dataset while utilizing only 70\% of the dataset and requiring far fewer trainable parameters than the current state-of-the-art model.
\end{itemize}

\section{Related work}
\label{sec:rel_work}

\subsection{Cytology classification}
Many cytology classification studies use pre-trained CNN models as feature extractors, typically fine-tuning only the classification layer added on top of the backbone~\cite{Jiang2023, Jiang2023_review}. For example, UÇA \textit{et al.}~\cite{UCAn2022} compared various CNN architectures, finding ResNet50 effective for body cavity cytologies. Yaman and Tuncer~\cite{Yaman2022} employed DarkNet19 to extract features from cervix cytological images, combined with traditional machine learning techniques. While CNNs remain prevalent, transformer-based models have recently gained attention for their ability to capture complex feature relationships~\cite{Jiang2023_review, Jiang2023}. Cai \textit{et al.}~\cite{cai2024hicervix} introduced HierSwin, a model based on Swin Transformer architecture that leverages hierarchical information, marking a shift toward more advanced transformer architectures in cytology.

\subsection{Foundation models} 
Certainly one of the most popular FM is the vision transformer introduced by Google~\cite{wu2020visual}. Beyond vision, FMs have been developed  for multiple modalities, including vision and language. Vision-language models, such as CLIP~\cite{Radford2021clip}, learn to align visual and textual embedding in a common embedding space using large-scale image-text datasets. Following the introduction of the first FMs, increasingly specialized datasets have been developed, leading to the emergence of domain-specific FMs. Examples include the vision-language models BiomedCLIP~\cite{Zhang2024biomedclip} and QUILT~\cite{Oluchi2023quilt}, focusing on biomedical and histopathological applications, respectively, as well as UNI~\cite{Chen2024uni}, a vision transformer also designed for histopathology. To the best of our knowledge, no FM has yet been introduced specifically for cytology.

\subsection{Parameter-efficient fine-tuning}
As fine-tuning these models requires extensive datasets and computational resources, parameter-efficient fine-tuning methods have gained popularity for training only a small subset of parameters, as highlighted in the review by Han \textit{et al.}~\cite{han2024}. Among these methods, Low Rank Adaptation (LoRA)~\cite{hu2021lora} updates models weight using low-rank matrices. Dutt \textit{et al.}~\cite{dutt2023parameter} evaluated various parameter-efficient fine-tuning methods on transformer-based models on different medical tasks and modalities. Their findings show that LoRA consistently outperforms full fine-tuning, especially in low-data scenarios, establishing it as the most effective parameter-efficient fine-tuning method across all tested datasets. Building on this, Zanella \& Ben Ayed~\cite{zanella2024cliplora} applied LoRA specifically to CLIP, demonstrating its ability to achieve strong generalization further supporting LoRA's applicability across diverse domains.

\section{Experimental setup}
\label{sec:method}
\subsection{Fine-tuning methods}
To adapt FMs to cytology classification, we compare two fine-tuning strategies. \\

\noindent \textbf{Linear classifier.}
A common fine-tuning approach is to leverage a pre-trained model $\theta$ to extract key features $z = f_{\theta}(x)$ from an input $x$ and to add a linear classifier on top. During training, only the weights of the linear layer are updated to tailor the model for the specific task, as described by the following equation for the forward pass: 
\begin{equation}
    \label{eq:lin_class}
    y = \textbf{W} z
\end{equation}
where $z \in \mathbb{R}^{n}$ and $\textbf{W} \in \mathbb{R}^{c \times n}$, with $n$ the number of extracted features by the model and $c$ the number of classes. 
\\

\noindent \textbf{LoRA.} 
LoRA~\cite{hu2021lora} is a parameter efficient fine-tuning method describing the update of pre-trained weights of model with low-rank matrices. Given the initial model weight $\textbf{W} \in \mathbb{R}^{m \times n}$, the weight update $\Delta \textbf{W} \in \mathbb{R}^{m \times n}$ is modeled as the multiplication of two matrices of small rank $r$, $\textbf{A} \in \mathbb{R}^{r \times n}$ and $\textbf{B} \in \mathbb{R}^{m \times r}$. This gives rise to a modification of the forward pass:
\begin{equation}
    \label{eq:forward}
    h = \textbf{W} x + \gamma \Delta \textbf{W} x = \textbf{W} x + \gamma \textbf{B} \textbf{A} x
\end{equation}
\noindent with $h$ a hidden state from an intermediate layer, $x$ the input of this layer and $\gamma$ a scaling factor. Matrix $\textbf{A}$ is randomly initialized using Kaiming initialization, whereas matrix $\textbf{B}$ is set to all zeros. 
The entries of these low-rank matrices serve as the trainable parameters, while the remaining weight matrices remain frozen during training, reducing the number of parameters to be trained. During inference, this number is unchanged, as weight update $\Delta \textbf{W}$ is added to the initial one. For instance, this can be applied to all or a subset of the attention matrices of the model (query, key, value and output)~\cite{zanella2024cliplora}.

\subsection{Datasets}
We evaluate the classification performance on four publicly-available cytological datasets.\\

The \textbf{Body Cavity Fluid Cytology}~\cite{kaggle2022bodycavity} (BCFC) dataset consists of isolated cell images from body fluid effusions. These cells are classified into two categories: benign or malignant. The \textbf{Mendeley LBC Cervical Cancer}~\cite{Hussain2020} (MLCC) dataset is composed of pap smears images of liquid based cytology, representing the four sub-categories of cervical lesions (malignant and pre-malignant) of the Bethesda System standard. \textbf{SIPaKMeD}~\cite{Plissiti2018sipakmed} is a cervical cancer dataset, containing manually cropped images of isolated cells coming from pap smear slides. They are divided into five classes of cells. \textbf{HiCervix}~\cite{cai2024hicervix} is a multi-center dataset of cervical cells extracted from whole slide images, resulting in 40,229 images. This makes HiCervix the largest publicly-available cervical cytology dataset. The classes are structured within a three-level hierarchical tree to capture detailed subtype information. In an effort to align with the state-of-the-art, we choose to focus on the third level of annotation, which consists of 25 classes.

\subsection{Models} 
We examine five FMs pre-trained on natural and medical images. Among these, three are vision-language models; note that in this study, we exclusively used their visual encoders. \\

\textbf{CLIP}~\cite{Radford2021clip}, a vision-language model trained on a dataset of 400 million image-text pairs sourced from the Internet, provides a strong foundation for general visual-language tasks. \textbf{ViT}~\cite{wu2020visual}, introduced by Google, is a vision transformer pre-trained on ImageNet-21k and fine-tuned on ImageNet 2012, making it highly effective for feature extraction in downstream tasks. \textbf{BiomedCLIP}~\cite{Zhang2024biomedclip} is a vision-language model pre-trained on PMC-15M, a dataset with a wide range of medical image modalities, making it particularly relevant for medical applications. \textbf{QUILT}~\cite{Oluchi2023quilt}, is a CLIP model fine-tuned on Quilt-1M, the largest vision-language dataset specifically focused on the histopathology. \textbf{UNI}~\cite{Chen2024uni} is a vision transformer more specific to medical as it is pre-trained on more than 100 millions histopathological images. It should be noted that, unlike previous models which use a ViT-B/16 backbone, UNI employs a ViT-L/16, composed of four times more parameters.

\section{Experiments}
\label{sec:exp} 
Results are averaged over 3 seeds. Top-1 accuracy is evaluated on the validation set to determine the best learning rate for each model-shots-dataset combination, and final reported performances are measured on the test set of each dataset. The batch size is set to 32.

\subsection{Experiment 1: Linear classifier}

\textbf{Experiment details.}
To assess the effectiveness of existing FMs in extracting discriminative features for cytological classification, we use the aforementioned models as feature extractors, keeping the backbone frozen and fine-tuning only a classification head (Eq. \eqref{eq:lin_class}). \\

\noindent \textbf{Results.} 
The results are summarized in Table \ref{tab:exp1}. On average, UNI outperforms the other models, ranking highest on three out of four datasets, while QUILT achieves better results on the first dataset. It's worth noting that UNI utilizes a ViT-L/16 backbone, resulting in four times more parameters compared to the other models. Excluding UNI, the second-best model on average across the four datasets is QUILT. The results demonstrate the effectiveness of histopathology-pre-trained FMs to extract discriminant features for cytology tasks. It should be noted that, for the first two datasets~\cite{kaggle2022bodycavity, Hussain2020}, UNI with a fine-tuned classifier achieves accuracies consistent with those reported in the literature.

\begin{table}[!t]
    \centering
    \caption{Mean accuracy of fine-tuned classifiers, evaluated on five models and four datasets.}
      \vspace{0.5mm}
    \resizebox{\linewidth}{!}{
    \begin{tabular}{cccccc}
        \toprule
 & \multicolumn{5}{c}{\textbf{Models}} \\
        \cmidrule{2-6}
\textbf{Datasets} & CLIP & QUILT & BiomedCLIP & UNI & ViT  \\
        \cmidrule{1-6}
BCFC & 94.74 {\footnotesize$\pm$0} & \textbf{100{\footnotesize$\pm$0}} & 96.17{\footnotesize$\pm$0} & 99.04{\footnotesize$\pm$0} & 98.72{\footnotesize$\pm$0.28}  \\
MLCC & 93.95 {\footnotesize$\pm$0.52} & 97.95{\footnotesize$\pm$0} & 95.1{\footnotesize$\pm$0.52} & \textbf{99.2{\footnotesize$\pm$0.4}} & 96.8{\footnotesize$\pm$0.2}  \\
SIPaKMed & 90.5 {\footnotesize$\pm$0.16} & 92{\footnotesize$\pm$0.57} & 87.62{\footnotesize$\pm$0.33} & \textbf{93.33{\footnotesize$\pm$0.16}} & 89.71{\footnotesize$\pm$0.49}  \\
HiCervix & 55.2{\footnotesize$\pm$0.2} & 54.64{\footnotesize$\pm$0.08} & 46.39{\footnotesize$\pm$0.1} & \textbf{64.07{\footnotesize$\pm$0.25}} & 57.64 {\footnotesize$\pm$0.14} \\
        \cmidrule{1-6}
\textbf{Average} & 83.59 & 86.15 & 81.32 & \textbf{88.91} & 85.72 \\
        \bottomrule
    \end{tabular}}
    \label{tab:exp1}
\end{table}

\subsection{Experiment 2:  LoRA few-shot adaptation}
\begin{figure}[!t]
  \centering
  \includegraphics[width=0.5\textwidth]{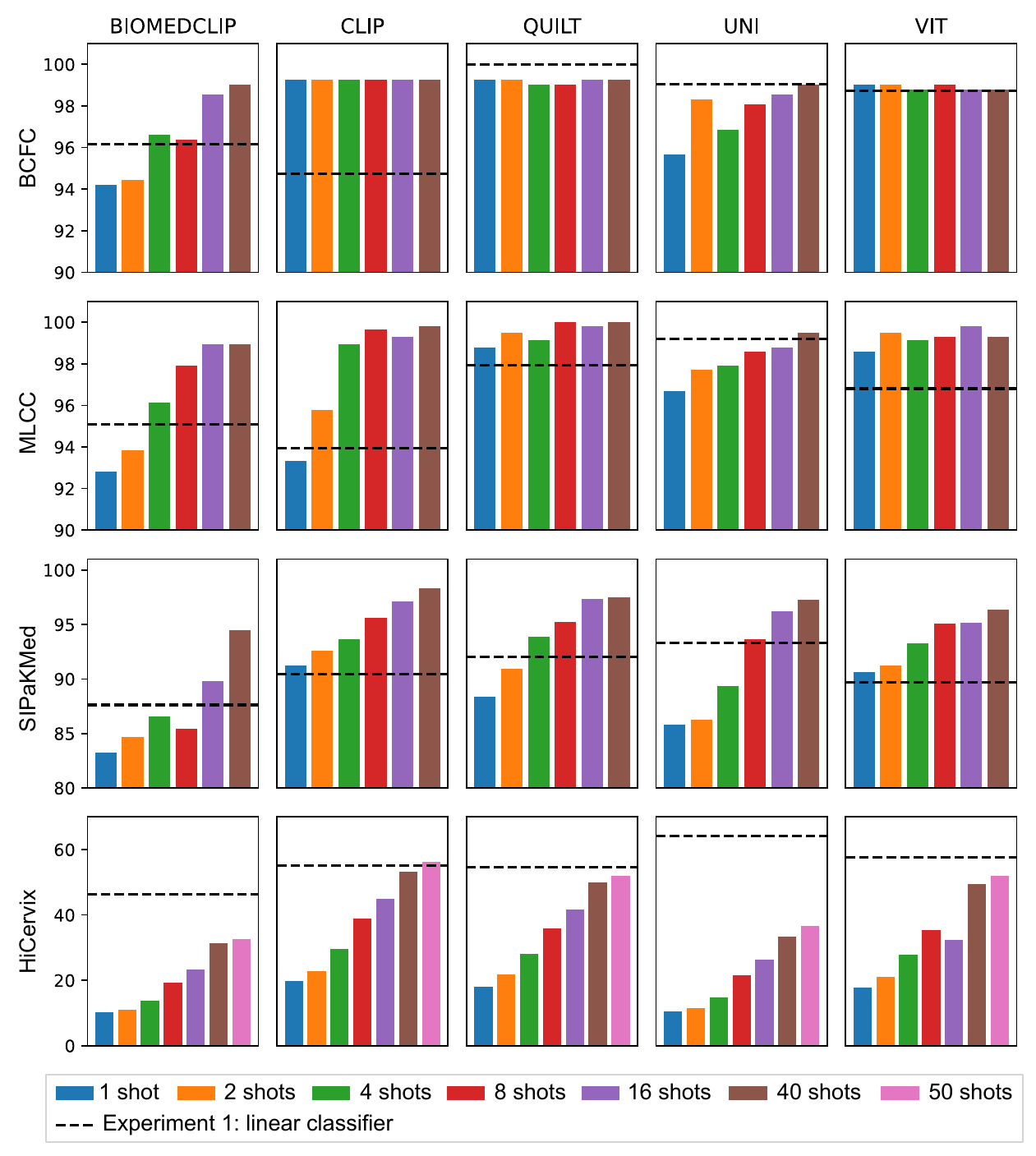}

  \caption{Mean accuracy of fine-tuned foundation models, evaluated on four datasets in a few-shot setting with a varying numbers of shots. The black horizontal line reports the mean accuracy from Experiment 1.}
  \label{fig:exp2}
\end{figure}

\textbf{Experiment details.}
The objective is to determine the gain of fine-tuning FMs beyond the classification head. Due to the large number of parameters in these models, we apply LoRA (Eq. \eqref{eq:forward}) to the query and value matrices of each attention block in the visual encoder. Based on the work of Zanella and Ben Ayed~\cite{zanella2024cliplora}, the rank of the LoRA matrices is set to 2. The model is trained in a few-shot setting with shots per class ranging from $1, 2, 4, 8, 16$, to $50$. For instance, 2 shots of the SIPaKMeD dataset results a total of 10 examples (two examples per class). \\

\noindent \textbf{Results.}
The results can be seen in Fig.\ref{fig:exp2}. CLIP fine-tuned with LoRA outperforms the linear classifier with only one or two samples per class on the first three datasets. For the HiCervix dataset, 50 shots are needed to achieve similar results, though this still represents only 4.4\% of the total data. Models pre-trained on medical or histopathological images require more data to reach comparable performance to those trained on more diverse datasets, emphasizing the strong generalization of models with broader pre-training when samples are limited. Overall, fine-tuning the entire backbone consistently yields higher accuracy than fine-tuning the classifier head alone, highlighting the advantage of backbone fine-tuning in data-limited scenarios.

\subsection{Experiment 3: Pushing model fine-tuning limits}

\textbf{Experiment details.}
Our goal is to determine whether the most generalizable model can achieve state-of-the-art performance using a smaller dataset proportion and fewer trainable parameters compared to the current state-of-the-art model. We focus on the HiCervix dataset, which is particularly large and presents a complex classification task due to its high number of classes. Based on Experiment 2, we select CLIP as the model with the best generalization capabilities. To further improve performance, we use a ViT-L/14 backbone and apply LoRA to the query, value, key, and output matrices with a rank of 16. It is important to note that, while ViT-L/14 is larger than the HierSwin model, the use of LoRA allows training only a portion of the parameters, reducing the number of trainable parameters to 3.1 million, which is 62 times fewer than current state-of-the-art model. Mean accuracy was evaluated across different dataset proportions, ranging from 5\% to 100\%, and compared to HierSwin’s state-of-the-art performance~\cite{cai2024hicervix}. 

\newpage

\begin{figure}[!t]
  \centering
\vspace{4mm}

  \includegraphics[width=0.5\textwidth]{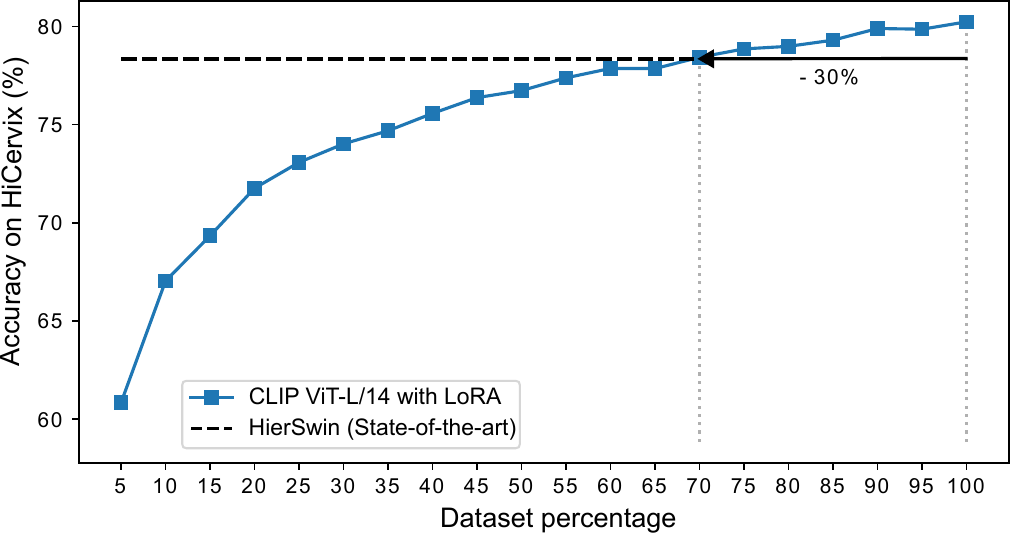}

  \caption{Mean accuracy of CLIP fine-tuned with LoRA and a ViT-L/14 backbone on different percentage of the HiCervix dataset. The black horizontal line represents the state-of-the-art accuracy (HierSwin).}
  \label{fig:exp3}
\end{figure}

\noindent \textbf{Results.} The results are shown in Fig.\ref{fig:exp3}. We observe that the model fine-tuned on 70\% of the HiCervix dataset achieves state-of-the-art performance, outperforming HierSwin in terms of both parameter efficiency and data usage. Specifically, this setup allows for high accuracy while updating significantly fewer parameters and utilizing fewer samples than required by HierSwin. When trained with the full dataset (100\%), the model’s accuracy further increases, reaching 80.23\%, suggesting that collecting more data could further improve model's performance. These findings highlight the model's ability to achieve competitive accuracy with a reduced training load.

\section{Conclusion}
\label{sec:conclu}

This study investigates the potential of foundation models for cytology. We show that fine-tuning with LoRA significantly improves performance over classifier-only fine-tuning, with particularly strong results in few-shot settings where labeled data is limited, and further gains when more data is available. Our findings suggest that while histology and cytology share visual features, models pre-trained on histology images do not necessarily generalize better for cytology. This is likely due to the absence of tissue structures in cytology, underscoring the need for cytology-specific adaptations in foundation models.\\

Future work could investigate the benefits of incorporating textual information through vision-language models, potentially enhancing model performance. Overall, we demonstrate that fine-tuning existing foundation models is deployable in cytology classification while remaining both practical and resource-efficient. Such adaptations could enhance the accuracy and accessibility of AI-driven cytology classification, supporting its growing role as a critical diagnostic tool.

\vfill
\pagebreak
\vfill

\section{Compliance with Ethical Standards}
This is a study for which no ethical approval was required.

\section{Acknowledgments}
\label{sec:acknowledgments}
M. Dausort and T. Godelaine are funded by the MedReSyst project, supported by FEDER and the Walloon Region.
M. Zanella is funded by the Walloon region under grant No. 2010235 (ARIAC by DIGITALWALLONIA4.AI).
Computational resources were made available on the Lucia infrastructure (Walloon Region grant n°1910247). 

\bibliographystyle{IEEEbib}

{\small\bibliography{biblio}}

\end{document}